# Unveiling the research landscape of Sustainable Development Goals and their inclusion in Higher Education Institutions and Research Centers: major trends in 2000-2017

Núria Bautista-Puig[1]; Ana Marta Aleixo[2]; Susana Leal[3]; Ulisses Azeiteiro[2]; Rodrigo Costas[4, 5]*

\* Corresponding author

Núria Bautista-Puig (ORCID: 0000-0003-2404-0683)
[1] INAECU (Research Institute for Higher Education and Science), University Carlos III of Madrid, Spain
E-mail: nbautist@bib.uc3m.es

Ana Marta Aleixo (ORCID: 0000-0002-4541-4560)
[2] Department of Biology & CESAM Centre for Environmental and Marine Studies, University of Aveiro, 3810-193, Aveiro, Portugal
E-mail: marta.aleixo@ua.pt

Susana Leal (ORCID: 0000-0002-8796-8289)
[3] Polytechnic Institute of Santarém & Life Quality Research Centre, Complexo Andaluz, Ap. 295, 2001-904 Santarém, Portugal
E:mail: Susana.leal@esg.ipsantarem.pt

Ulisses Azeiteiro (ORCID: 0000-0002-5252-1700)
[2] Department of Biology & CESAM Centre for Environmental and Marine Studies, University of Aveiro, 3810-193, Aveiro, Portugal
Email: ulisses@ua.pt

Rodrigo Costas (ORCID: 0000-0002-7465-6462)
[4] Centre for Science and Technology Studies (CWTS), Leiden University, Leiden, The Netherlands.
[5] DST-NRF Centre of Excellence in Scientometrics and Science, Technology and Innovation Policy, Stellenbosch University, Stellenbosch, South Africa
E-mail: rcostas@cwts.leidenuniv.nl

**Abstract**

Sustainable Development Goals are the blueprint to achieve a better and more sustainable future for society. Its legacy is linked with the Millennium Development Goals, set up in 2000. A bibliometric analysis was conducted to 1) measure 'core' research output from 2000-2017, with the aim to map the global research of sustainability goals, 2) describe thematic specialization based on keywords co-occurrence analysis and strongest citation burst, 3) present a methodology to classify scientific output (based on an ad-hoc glossary) and assess SDGs interconnections.

Sustainability goals publications (core+expand based on direct citations) were identified in-house CWTS Web of Science by using search terms in titles, abstracts, and keywords. 25,299 bibliographic records were analyzed, from which 21,653 (85.59%) are from HEIs and research centres (RC). The purpose of this paper is to analyze the role of these organizations in sustainability research. The findings reveal the increasing participation of these organizations in this research (660 institutions in 2000-2005 to 1744 institutions involved in 2012-2017). In terms of specialization, some institutions present a higher production and specialization on the topic (e.g., London School of Hygiene & Tropical Medicine and World Health Organization); however, others present less production but higher specialization (e.g., Stockholm Environment Institute). Regarding the topics, health (especially in developing countries), women and socio-economic aspects are the most prominent ones. Moreover, it is observed the interlinked nature of SDGs between some SDGs in research output (e.g., SDG11 and SDG3). This study provides important orientation for HEIs and RCs in terms of Research,

Development and Innovation (R&D+i) to respond to major societal challenges and could be useful for the policymakers in order to promote the research agenda on this topic.

1. **Introduction**

*1.1. Increasing awareness-building in Sustainable Development Goals*

Sustainable goals have emerged as a global strategy to solve critical world problems, as a result of a global environmental concern started in the 1970s. The origin of the Sustainable Development (SD) movement can be traced back to its most-recognized milestone in 1987: the definition of SD[1] in the Brundtland Report. Afterwards, different summits and conferences where held in which sustainability and SD were the core discussion (e.g., Earth Summit in Rio de Janeiro). In 2000, the Millennium Summit led to the Millennium Declaration with the creation of the eight Millennium Development Goals (MDGs)to be achieved by 2015. These goals tackled topics such as extreme poverty and hunger, child mortality or maternal health[2]. They represented an unprecedented effort to tackle the needs of the world's poorest countries. However, these goals have been criticized for not being adequately aligned 'with human rights standards and principles' (International Human Rights Instruments, 2008), and because they were relevant and focused only on developing countries (Fukuda-Parr, 2016). Another criticism of the MDGs is that they had unsuccessful effects in some regions like Africa (Easterly, 2009). However, although not all goals have been achieved by 2015, some progress has been acknowledged (United Nations, 2015a). For instance, 'the number of people living in extreme poverty has declined by more than half since 1990 and the literacy rate among youth aged 15 to 24 has increased globally, from 83% in 1990 to 91% in 2015' (Ki-Moon, 2013).

In 2012, the Conference Rio+20 adopted a ten-year plan called Agenda 2030 (2015-2030) for sustained economic growth, social development, and environmental protection (United Nations, 2015b). As a result, they established 17 Sustainable Development Goals (SDGs) with their related indicators, with a deadline in 2030. The agenda has 169 targets and various indicators for monitoring their achievement. The topics of these goals cover five critical areas (the 5 P's): People, Planet, Prosperity, Peace, and Partnership (United Nations, 2015b). The motto of these goals is 'no one left behind': that is, all countries need to be involved. Monitoring has also become a key issue for SDGs in comparison with MDGs and, since the launch of SDGs, an SDG Index[3] has been conducted with the aim to evaluate the achievement of each goal by countries. The index allows identifying priorities for action, support discussions, debates or identify gaps in the data. A preliminary set of 330 indicators was introduced in March 2015 (Hák, Janoušková, & Moldan, 2016), but only 232 indicators were adopted. Different from the MDGs, in which the indicators were decided on an internal basis, the SDG indicators are based on public consultation that was led by the Open Working Group established in 2013. Moreover, the indicators 'come from a mix of official and non-official data sources', subjected to an extensive and rigorous data validation process (e.g., the World Bank, the Organisation for Economic Cooperation and Development, among others) (Stiftung & SDSN, 2019). The number of countries that participated in this monitoring platform has also increased over time: from 34 in 2015 in the first edition (Kroll, 2015) to 162 in 2019.

*1.2. SDGs and their relation with HEIs and RCs*

MDGs and SDGs appeared as a result of the interest and commitment of the different world countries towards sustained growth. As Caiado et al. (2018) stated, 'The SDG agenda calls for a global partnership – at all levels – between all countries and stakeholders who need to work together to achieve the goals and targets, including a broad spectrum of actions such as multinational businesses, local governments, regional and international bodies, and civil societal organizations'. In this regard, Higher Education Institutions (HEIs) and Research Centres (RCs) should play an active and central point in promoting and participating in these new goals. In the past, HEIs played a role in 'transforming societies and serving the greater public good, so there is a societal need for universities to assume responsibility for contributing to SD' (Waas et al., 2010) and 'they should be leaders in the search for solutions and alternatives to current environmental problems and agents of change' (Hesselbarth and Schaltegger, 2014). For Bizerril et al. (2018), the knowledge of sustainability in HEIs should be encouraged worldwide and especially those located in regions with serious social and environmental challenges. In this sense, researchers must discuss how to cooperate and to share knowledge for a sustainable society, and the network could respond to HEIs sustainability through cooperation. According to Lozano et al. (2015), HEIs (and in extension, RCs) could tackle the SD by the following aspects: 1) Institutional framework (i.e., the HEIs commitment with policies); 2) Campus operations; 3) Education; 4) Research; 5) Outreach and collaboration (e.g., exchange programmes for students in the field of SD); 6) SD through on-campus experiences; 7) Assessment and reporting. Despite all these pillars, as Caeiro et al. (2013) study stated, only a few institutions follow a holistic implementation, in which SD is applied in all pillars via its inclusion in social, economic and environmental pillars.

---

[1] SD was defined as a "kind of development that meets the needs of the present without compromising the ability of future generations to meet their own needs" (United Nations, 1987).
[2] Information of the MDGs available at the following link: https://www.who.int/topics/millennium_development_goals/about/en/ accessed 30 December 2019.
[3] SDG Index available at: http://sdgindex.org/- accessed 30 December 2019.

As part of its contribution to the achievement of SDGs, research is one of the most relevant dimensions. According to Tatalović and Antony (2010) science did not factor strongly in discussions how to achieve the goals during the MDGs. Leal Filho et al. (2017) referring to SDGs as an opportunity for SD research and as a result to solve SDGs. For Leal Filho et al. (2018), development goals are an opportunity to encourage sustainability research through interdisciplinary and transdisciplinary research. Several authors reaffirm the importance of research to achieve SDGs (Wuelser and Pohl, 2016), namely as a way to solve concrete social problems, and science of sustainability could support the transaction for sustainability.

Bibliometrics is a research area focused on examining the scientific activity in a given subject area, institution, or country. Bibliometric analyses offer a powerful way to generate a global picture of research in a particular area. There are different bibliometric studies that analyze sustainability or SD (Pulgarin et al. 2005; Hassan, Haddawy and Zhu, 2014; Ramírez, 2016; Olawumi and Chan, 2018;) or sustainability science (Kajikawa et al., 2014; Nučič, 2012; Schoolman et al., 2012). Other studies focus on analyzing the output of sustainability in higher education (Bizerril, 2018; Veiga-Ávila et al., 2018; Alejandro Cruz et al., 2019; Hallinger and Chatpinyakoop, 2019). However, few studies have specifically analyzed scientific output on SDGs. As a preliminary step for this study, Bautista-Puig and Mauleón (2019a) analyzed the *core* of scientific production on the MDGs and the SDGs (n=4,532) in addition to the interrelations between different SDGs. Nakamura (2019) analyzed 2,800 publications (with an expansion of 10,300), as well as its SDG topic map. Regarding research output on SDGs at HEIs, the Aurora project[4] analyzes the scientific production through a bibliometric analysis. Other studies analyze the interrelations among SDGs (Griggs et al., 2017, Le Blanc, 2015). However, to the best of our knowledge, no other study has approached the study of the SDG relation in scientific output from a bibliometric point of view.

*1.3. Objectives*

Bearing in mind the importance of SDGs, our main objective is to study the development of research in MDGs and SDGs (henceforth M&SDGs) in the HEIs and RCs from 2000 through 2017, by analyzing the institutions involved in their activity, as well as the topics related to this research. This bibliometric analysis was guided by two research questions:
- RQ1: How the research on M&SDGs carried out by HEIs has developed over time? This question seeks to understand when the sustainability concept appears on the scientific literature, who are the main actors (institutions and countries) and how this topic has evolved over time.
- RQ2: What are the thematic relationships between SDGs? This question analyzes the main research topics studied in the scientific literature and the interrelations among 17 SDGs based on an *ad-hoc* glossary (Bautista Puig, 2019b).

The rest of the article is organized as follows. Section 2 covers the research methodology. Section 3 introduces the main results and discussions, providing answers to the research questions. Finally, section 4 presents the main conclusions and suggestions for future research.

**2. Method**

The following methodological steps were followed: (i) Formulation of a search strategy to identify the *core* M&SDGs literature; (ii) Expansion of the dataset based on direct citations (cited and citing publications); (iii) Data collection and information processing; (iv) Development of bibliometric indicators.

In the first step, we designed a search strategy composed by M&SDGs keywords in title, abstract and keywords (author and paper keywords)[5]. The search strategy above was run in the in-house CWTS WoS database (from 2000 to 2017). A total of 4,685 publications

---

[4] Aurora project information available at: https://aurora-network.global/project/sdg-analysis-bibliometrics-relevance/ - accessed 10 February 2019
[5] Search query used in this study: TS="Millennium Development Goal*" OR TS="Millennium Goal*" OR TS="Sustainable Development Goal*"

were collected. These are considered as the *core* set of publications of this research. In a second step, the set of their direct citations (DC), considering both cited (n=59,180) and citing (n=74,859) publications were collected, resulting in a final set of distinct publications "M&SDGs Expansion" (n=129,379). In a following step, a total of 25,299 publications between 2000 and 2017 were considered, of which only publications with at least one affiliation of a HEIs or a RCs were finally selected (n=21,653)[6], all restricted to the period 2000 to 2017[7] (Figure 1), conforming the final dataset of analysis ("Dataset of M&SDGs"). The harmonization of the affiliations was based in the in-house CWTS database (Waltman et al., 2012).

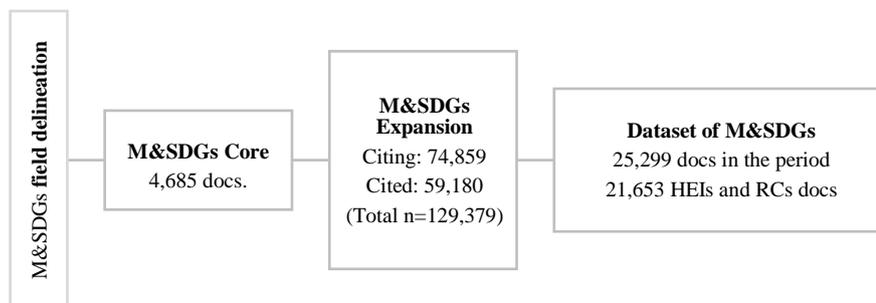

**Figure** 1. Methodological workflow for delineating M&SDGs on this study and creating the final dataset

An important conceptual difficulty in the definition of what production can be related to M&SDGs research is the discrepancy between esearch *related to M&SDGs* and *research on M&SDGs*. The first approach comprises research that is *related to* concepts, issues or ideas related to the SDGs but without necessarily a direct linkage to the M&SDGs core (e.g. an institution doing research related to malaria prior to the official launch of the SDGs). The second approach would consider research directly focusing *on* the concepts, issues or ideas of the M&SDGs. In this work we opt for using the second approach. Thus, we consider that research on M&SDGs is either mentioning the concept (i.e. core research) or at a minimum being cited by or citing the core research. As such, it can be argued that this work focuses on the 'discourse of sustainable goals', about how this topic has been constructed in the research by HEIs and RC, providing a focused view on scientific research with a stronger alignment with the M&SDGs philosophy and aims.

The following indicators were analyzed for the final dataset:

(i) Research patterns

- **Yearly trend** in scientific output in M&SDGs overall and by HEIs and RCs. A trend analysis of 6-year blocks is considered .

- **Cumulative Average Growth Rate (CAGR).**

- **Output by institutions and countries**: absolute values and "**Activity Index**" (AI) of their M&SDGs research (Eq. A.1.). The AI was proposed by Frame (1977) and it is used to analyse the degree of relative specialization of an actor (institution or country) in a research field. The indicator represents the percentage of contribution of each country to the total WoS production compared to the percentage of contribution in the analyzed topic.

(ii) Subject specialization.

- **Co-occurrence map**[8] **based on keywords using the VOSviewer tool**[9] to identify thematic clusters within the scientific landscape. Regarding the clustering, VOSviewer applies its own algorithm based on modularity optimization (Van Eck and Waltman, 2017). The following indicators has been analyzed:

**Table 1.** List of indicators analyzed

| Indicator | Description |
|---|---|
| Label | Name of the cluster considering its terms |
| #nodes | Number of nodes (terms) within the cluster |
| Core papers (and %) | Number and percentage of *core* publications |
| #link$_{avg}$ | Links per paper |
| #year$_{avg}$ | Average year of the publications in which a term occurs. |

---

[6] Documents excluded from the M&SDGs Expanded set are publications outside the period considered (2000-2017), but also publications without HEIs or RCs affiliations.
[7] The period corresponds with the launch of the MDGs in 2000.
[8] Co-citation is defined as the frequency with which two publications are cited together by other publications.
[9] VOSviewer is a software tool for constructing and visualizing bibliometric networks. These networks may include nodes of journals, researchers, or individual publications, and they can be constructed based on citation, bibliographic coupling, co-citation, or co-authorship relations. Additionally, also offers text mining functionality that can be used to construct and visualize co-occurrence networks of terms extracted from a research dataset (https://www.vosviewer.com/ - accessed 30 December 2019).

- **Keywords bursts citation**. Burst is a concept associated with a change of a variable's value in a relatively short time. Those sudden increases in the usage frequency of keywords (i.e. burst strength) in order to determine the *hotness* of a topic were identified using Kleinberg's algorithm (Kleinberg, 2003).
- **Classification of the scientific production into the SDGs**. In order to study the semantic relations between the different SDGs (in terms of SDGs sharing similar keywords across publications), the individual publications were classified into the different SDGs. To classify the publications into individual SDGS, an *ad-hoc* ontology (Bautista-Puig, 2019b) with 4,122 terms has been applied. Publications were classified in different individual SDGs based on the linkage between the keywords in the publications and the ontology, allowing publications to be classified in more than one SDG when their keywords would point to different SDGs. A total of 20,749 (82.01%) publications were finally classified in at least one of the 17 SDGs. This includes keywords related to each SDG based on the United Nations-Description (e.g., 'poverty' was classified into 'SDG1-No poverty', 'sanitation' into 'SDG6- clean water and sanitation')[10], as well as a manual-supervision of the keywords located on the *core* and its consequent extension.

## 3. Results

In this section the main results of the paper are presented in relationship to the main research questions formulated above.

### 3.1. RQ1. How the research on M&SDGs carried out by HEIs has developed over time?

*3.1.1. Research output and main actors*

This section analyzes the research output collected, as well as the main actors and their specialization. Figure 2 presents the evolution of the scientific output of development goals produced by HEIs and RCs. During the period of study, 21,653 publications produced by HEIs and RCs were detected, which represent 85.59% of the publications retrieved by the search strategy in the same period ($n = 25,299$) (Figure 2). Other types of organizations have participated in less than 7% of the publications (e.g. Hospitals appear as authors in 6%, Governmental organizations in just 3%). This reinforces the idea that HEIs and RCs play a key role in the production of M&SDGs scientific publications. This evolution shows a growing tendency, with an overall growth of 828.65% over the period and a CAGR of 14.01%. It is remarkable that since the launch of the SDGs in 2015, the scientific production represented more than 31.6% of the overall output, thus suggesting a strong concentration of output in M&SDGs in the period after 2015.

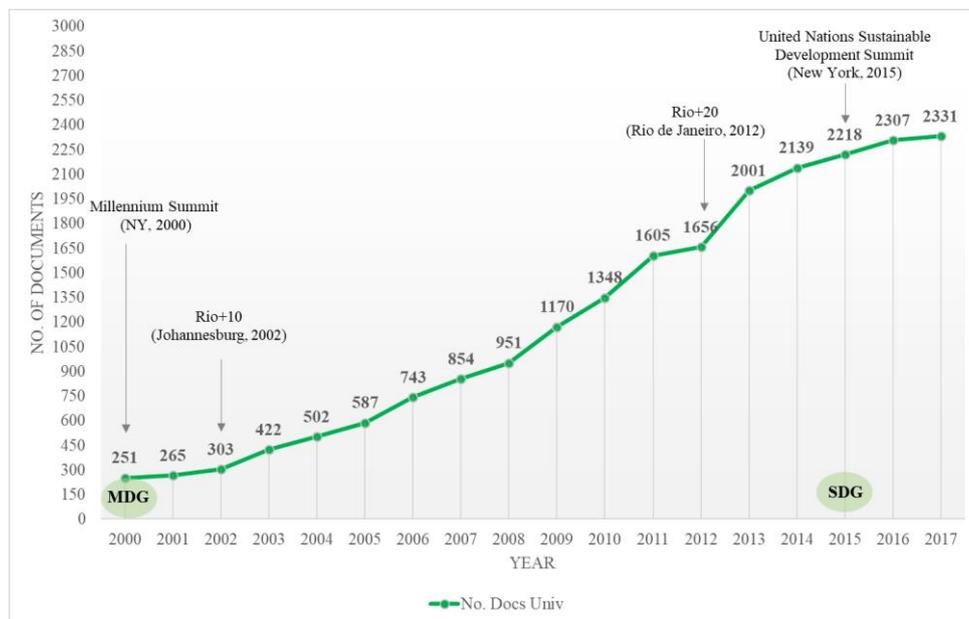

**Figure 2.** Yearly output of the scientific production of the HEIs and RCs (2000–2017).

A total of 1,968 organizations were identified in the affiliations of these publications. The most productive institution was the London School of Hygiene and Tropical Medicine, with 1,963 publications (9.07%), followed by the World Health Organization (WHO), with 1,675 (7.74%), Johns Hopkins University with 1,324 (6.11%) and Harvard University with 1,079 (4.98%). However, when looking at the six-year blocks as in Table A.1, different tendencies are shown over time. In the first 6-year (2000–2005) the number of publications was 2,330 produced by 660 organizations identified. The most productive organizations in this period were the WHO, with 292 publications (12.53%), followed by the London School of Hygiene and Tropical Medicine with 272 (11.67%) and the Johns Hopkins University with 157 (6.74%). In the second period (2006–2011), a total of 6,671 publications were produced by 1,244 organizations. During this period, the London School of Hygiene and Tropical Medicine led the ranking with 682 publications (10.22%), followed by the WHO with 580 (8.69%) and the Johns Hopkins University with 439 (6.58%). In the third period (2012–2017), a total of 12,652 publications, produced by 1,744 organizations, were identified. The same ranking of organizations as in the previous period is also found: The London School of Hygiene and Tropical Medicine leads with 1,009 publications (7.98%), followed by the WHO with 803 publications (6.35%) and the Johns Hopkins University with 728 (5.75%). Among the more productive HEIs there are only five institutions from developing countries: two form South Africa (the University of Cape Town and the University of the Witwatersrand), one from Uganda (Makerere University), one from Pakistan (Aga Khan University) and one from Brazil (the Federal University of Pelotas).

---

[10] Information available at: https://www.un.org/sustainabledevelopment/sustainable-development-goals/ - accessed 30 December 2019.

Figure 3 shows a scatter plot of the relation between the institutions with a higher scientific production on SDGs (P(M&SDGs)) and their AI around M&SDGs research (AI(M&SDG)). The size of the bubbles indicates the number of publications in WoS of each institution (only institutions with more than 50 are included in the Figure). Overall, the most productive institutions present a lower AI (e.g. the Johns Hopkins University and Harvard University with P(M&SDG)=1,324 publications and P(M&SDG)=1,024, respectively, have an AI of 8.70 and 3.89 respectively). The WHO (P(M&SDG)= 1,675) and the London School of Hygiene and Tropical Medicine (P(M&SDG)= 1,963) present a high AI of more than 88% each. Among the institutions with the larges AI values we find other institutions such as the Stockholm Environment Institute (AI 190.47), Aga Khan University (AI 141.06) or the International Centre for Diarrhoeal Disease Research, Bangladesh (AI 132.55) (Figure 3).

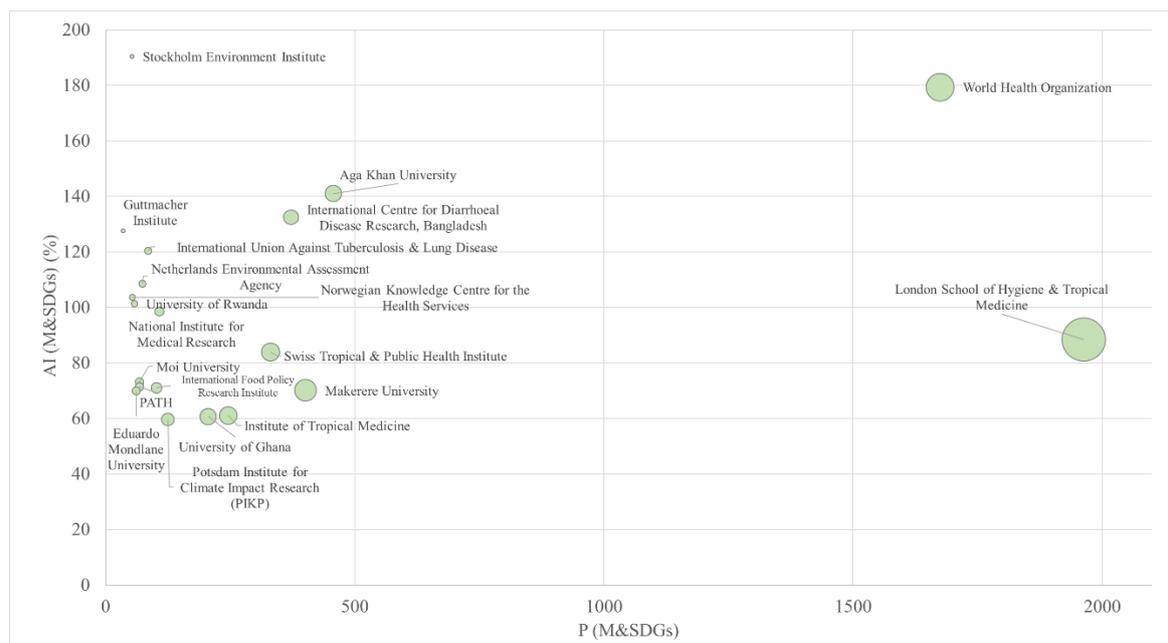

**Figure 3.** Scatter plot of the Top 20 organizations ranked by AI (>50 docs.).

A map is drawn in order to show the geographical distribution of M&SDGs publications (Figure 4). The most productive countries during the whole period were the United States (8,473 publications, 39.13%), followed by the United Kingdom (6,053 publications, 27.95%), Switzerland (2,232 publications, 10.31%), Australia (1,959 publications, 9.05%) and Canada (1,757 publications, 8.11%). By periods, in the first one (2000-2005) at total of 67 countries produced at least one publication on M&SDGs research increasing to 86 countries in the second period, and to 95 countries in the third period, with the same set of countries mentioned above as the most productive in each period (Figure 4). From the point of view of the specialization (measured by the AI), African and Asian countries exhibit a stronger specialization in M&SDGs research compared to countries from other regions. Uganda is leading the specialization in the whole period (29 publications and AI of 24% in the first period; 107 and AI 32.60 in the second period and 265 pubs and AI of 43.130 in the third period). In the first period (2000-2005) also Tanzania (21 publications, AI=13.61), Bangladesh (*n*=34, AI=13.35), Pakistan (*n*=31, AI 6.9), South Africa (*n*=106, AI=4.20) and Switzerland (*n*=327, AI=3.60) stand out in specialization. In the second period, after Uganda, the following countries are found: Bangladesh (*n*=131, AI=25.53), Zimbabwe (*n*=34, AI=25.47), Ghana (*n*=55,

AI=22.62), Jamaica (*n*=21, AI=18.47) and Tanzania (*n*=50, AI=16.16). In the third period, Rwanda is in the second position (*n*=50, AI=40.06), Ghana (*n*=215, AI=38.59), Mozambique (*n*=45, AI 31.46) and Ethiopia (*n*=149, AI 23.05).

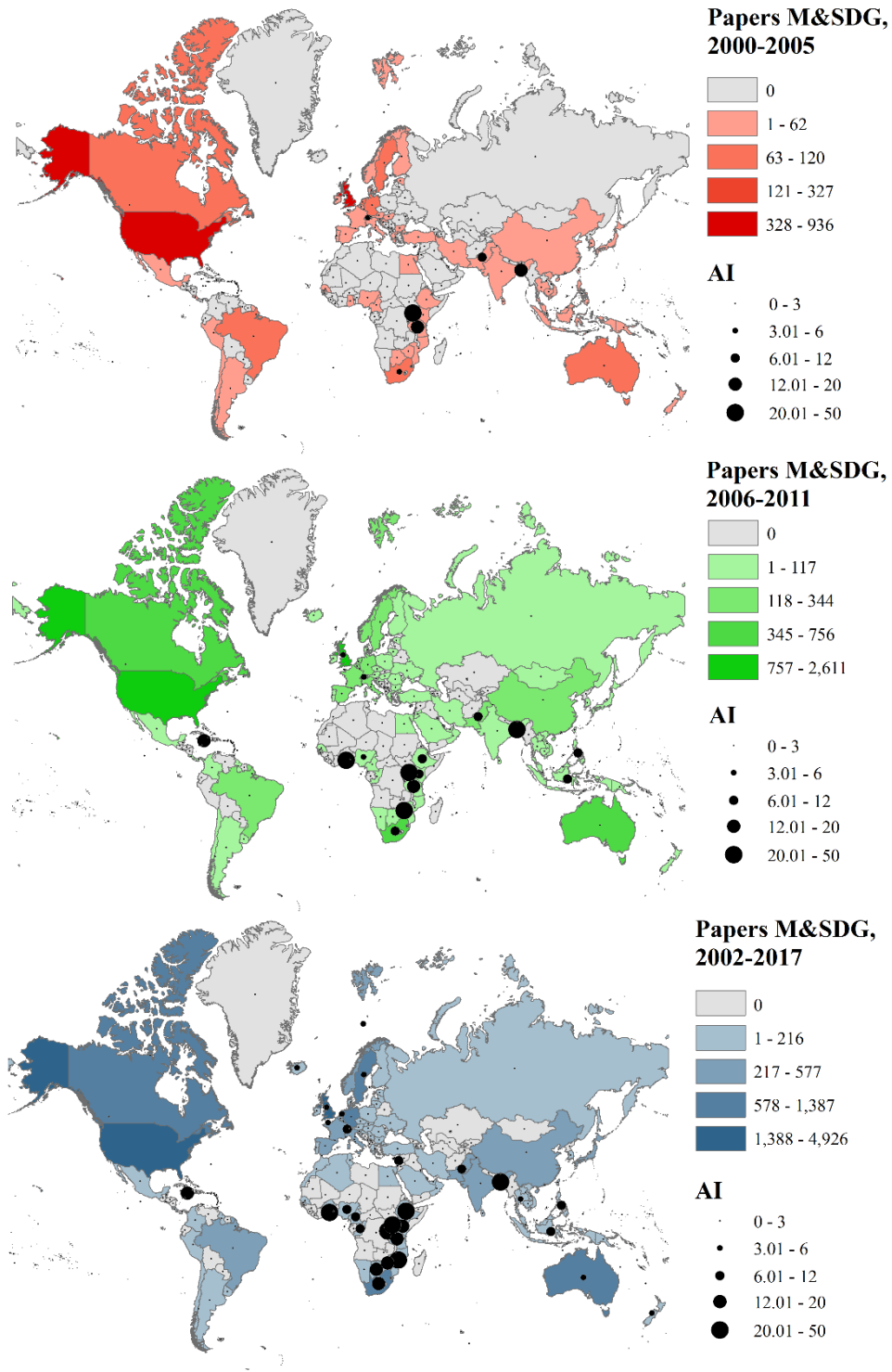

**Figure 4.** Geographic distribution of scientific publications and AI (countries with >20).

## 3.2. RQ2: What are the thematic relationships between SDGs?

*3.2.1. Keyword co-occurrence analysis*

To reveal the main topics of the M&SDGs research Figure 5 shows a keyword co-occurrence-based clustering[11]. Keywords (nodes) in VOSViewer maps are located in a way that the distance between them is related to their co-occurrence frequency. Terms located closely in the map means that they tend to appear together in the titles and abstracts of the papers, and therefore it can be argued that they are thematically connected. The following five clusters were identified: Cluster #1, with terms related to the millennium development goals inheritance and policy framework; Cluster #2 with terms about maternal mortality and care; Cluster #3 with terms related to the health systems ("diagnosis", "treatment"); Cluster #4 with terms about the African health ecosystem, and Cluster #5 including terms related to the developing countries landscape (health, community, water, and so on). Table 2 summarizes the main information of each cluster (number of nodes, core papers, average year, average links and the most frequent keywords). It can be observed that cluster 1 is the largest in terms of publications, followed by cluster 2. The number of links per paper (#link$_{avg}$) is higher in cluster #2 and cluster #3, both related to health issues, suggesting a stronger connection between these two clusters. In most clusters, the average year (#year$_{avg}$) is 2012, suggesting that an important share of the output has been developed in the most recent years of the study, which is backed up by the growing M&SDGs output over time discussed above. The percentage of core publications (i.e. directly referring to M&SDGs) on each cluster are indicated in the column "% core papers", showing that clusters #1 and #2 (with 45.40% and 37.55% of core publications respectively), are clusters with a stronger conceptual proximity with the M&SDGs core ideas and aims, while the other clusters have a more indirect relationship with these core ideas.

**Table 2.** Summary of five thematic clusters

| Cluster | Label | #nodes | Core papers | % core papers | #link$_{avg}$ | #year$_{avg}$ | Most-frequent terms and frequency |
|---|---|---|---|---|---|---|---|
| #1 | MDGs inheritance and policy framework | 160 | 2,127 | 45.40 | 212.72 | 2012.17 | management (967); policy (695); poverty (658); millennium development goals (640); climate-change (514) |
| #2 | Maternal mortality and care | 152 | 1,759 | 37.55 | 258.78 | 2012.53 | care (1,593); countries (1,351); interventions (933); services (883); maternal mortality (877) |
| #3 | Health systems: diagnosis, treatment… | 149 | 1,426 | 30.44 | 231.79 | 2012.15 | mortality (2,440); health (2,073); systematic analysis (774); randomized controlled-trial (768); risk-factors (730) |
| #4 | Africa health ecosystem | 82 | 1,000 | 21.34 | 218.74 | 2011.17 | Africa (1,257); sub-Saharan Africa (1,137), impact (1,128); South-Africa (741); malaria (563) |
| #5 | Developing countries and water sanitation | 24 | 531 | 11.33 | 219.57 | 2012.12 | developing countries (1,901); community (457); sanitation (373); water (342); diarrhea (307); drinking-water (215) |

---

[11] The parameters are detailed below: LingLog Modularity normalization method, 566 items, link of 65,446, link strength of 298,485 and repulsion, resolution and minimum cluster size with a value of 1.

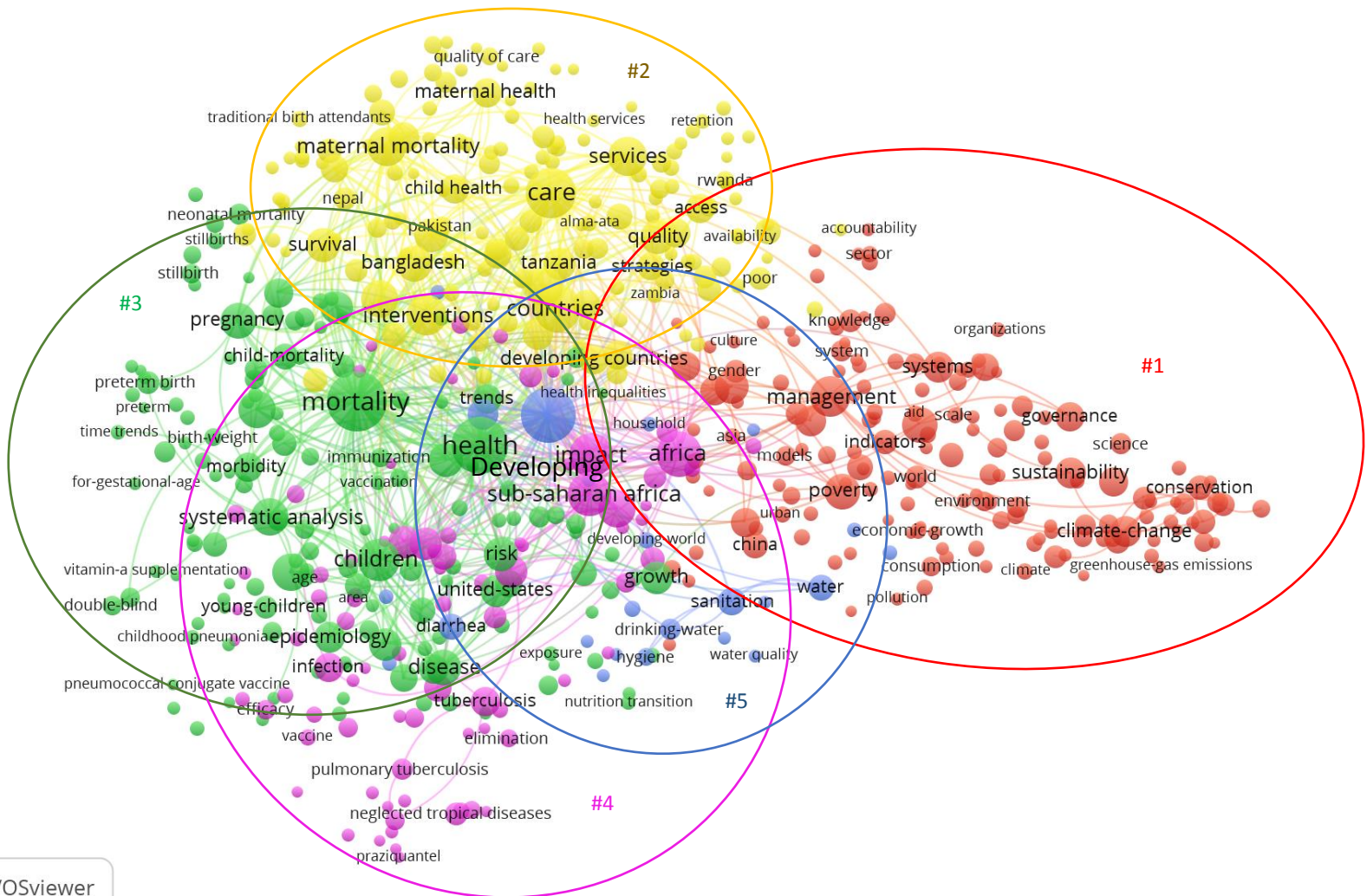

**Figure 5**. Co-ocurrence map (<50 keywords) of scientific research.

*3.2.2. Burst analysis of keywords*

In this section, burst detection for keywords in M&SDGs publications is performed in order to show what terms have more rapidly increased attention in citations accumulation. We have identified at least 60 different bursting keywords during the period. Table A.2 lists the 60 keywords with the strongest citation bursts, along with their strength and time span. The term "middle income country" has the strongest citation burst with a burst strength of 75.13, followed by "tuberculosis" with 66.52 and "maternal health" with 64.98. Some keywords have only been "bursting" at the very beginning of the period (e.g., "low birth weight", 2000–2003; "economic growth", 2000–2001; and "rural Bangladesh", 2000–2001). However, in the more recent years, strongly bursting citation keywords include "newborn" (16.65, time span of 2015–2017), "middle income country" (75.13, 2014–2017), "maternal health" (64.98, 2014–2017) and "delivery" (36.38, 2014–2017).

*3.2.3. Individual SDGs analysis*

*Publication prevalence*

The following SDGs were most prevalently represented in the publications (Figure A1): SDG3: Good Health and well-being, with 15,963 papers (76.93%); followed by SDG16: Peace, justice and strong institutions, with 11,658 (56.19%); SDG11: Sustainable cities and communities, with 9,541 publications (45.98%); and SDG10: Reduce inequalities, with 6,115 publications (29.47%). On the other hand, the least represented SDGs are: SDG 12: Responsible production and consumption, 939 papers (4.51%); and SDG7: Affordable and clean energy, with 1,095 (5.26%).

*Geographic distribution*

Table 3 shows two different perspectives on the production of publications across continents related to their contribution to the research of each individual SDG. In the left table, the contribution of each continent to each SDGs is presented (i.e. it must be read row-wise); while the table on the right depicts the share of each continent across the different individual SDGs (i.e. it must be read column-wise). Publications are assigned to each continent based on the affiliation of the first-author of the paper. The results of the left table show that all goals have higher production in Europe and North America. Considering all M&SDGs research, it can be observed that in Europe the largest percentage of output is in SDG13: climate action (46.23%), followed by SDG12: Responsible consumption and production (44.85%) and SDG15: life on land (44.28%). In America, the largest is SDG2: zero hunger (37.60%), followed by SDG5: gender equality (15.50%), and SDG3: good health (13.32%). In Africa, the highest production is in SDG5: gender equality (15.50%); SDG4: quality education (14.27%); and SDG11, (13.73%). In Asia, it is in SDG17: partnership for the goals (13.97%); SDG4 (13.33%), and SDG5 (13.31%). Finally, in Oceania, the higher production of these institutions is in SDG13 (8.47%); SDG12 (7.24%) and SDG15 (6.81%). From a global perspective, if we consider the distribution of the publications on each goal by continent to determine their profile (Table 3, right), the approach of the different SDGs exhibit more similar patterns, although some SDGs—such as good health (SDG3); peace, justice (SDG16); and reducing inequalities (SDG10)—stand out from the others.

**Table 3**. Contribution of each continent to each SDGs (left) and profile by continent (right)

| | Africa | America | Asia | Europe | Oceania | | Africa | America | Asia | Europe | Oceania |
|---|---|---|---|---|---|---|---|---|---|---|---|
| SDG1 | 10.23 | 33.23 | 10.17 | 40.43 | 5.94 | SDG1 | 1.82 | 1.97 | 1.81 | 2.33 | 2.12 |
| SDG2 | 9.42 | 37.60 | 11.52 | 35.00 | 6.46 | SDG2 | 3.55 | 4.70 | 4.32 | 4.25 | 4.87 |
| SDG3 | 13.32 | 36.70 | 11.58 | 33.15 | 5.24 | SDG3 | 23.95 | 21.93 | 20.73 | 19.23 | 18.86 |
| SDG4 | 14.27 | 34.94 | 13.33 | 32.09 | 5.36 | SDG4 | 5.30 | 4.32 | 4.93 | 3.85 | 3.99 |
| SDG5 | 15.50 | 37.43 | 13.31 | 28.24 | 5.52 | SDG5 | 5.25 | 4.21 | 4.49 | 3.08 | 3.74 |
| SDG6 | 10.80 | 33.71 | 10.29 | 40.87 | 4.32 | SDG6 | 2.87 | 2.98 | 2.73 | 3.51 | 2.30 |
| SDG7 | 4.58 | 32.21 | 13.07 | 43.70 | 6.44 | SDG7 | 0.55 | 1.29 | 1.57 | 1.70 | 1.55 |
| SDG8 | 10.63 | 34.98 | 10.53 | 37.45 | 6.41 | SDG8 | 4.99 | 5.46 | 4.92 | 5.67 | 6.01 |
| SDG9 | 9.43 | 36.05 | 10.49 | 37.60 | 6.43 | SDG9 | 2.79 | 3.55 | 3.10 | 3.59 | 3.81 |
| SDG10 | 9.71 | 34.46 | 10.42 | 39.59 | 5.82 | SDG10 | 6.69 | 7.89 | 7.14 | 8.80 | 8.02 |
| SDG11 | 13.73 | 34.79 | 13.07 | 33.07 | 5.35 | SDG11 | 14.75 | 12.42 | 13.99 | 11.46 | 11.49 |
| SDG12 | 6.03 | 30.26 | 11.62 | 44.85 | 7.24 | SDG12 | 0.62 | 1.03 | 1.19 | 1.49 | 1.49 |
| SDG13 | 4.59 | 32.06 | 8.65 | 46.24 | 8.47 | SDG13 | 0.88 | 2.04 | 1.65 | 2.86 | 3.24 |
| SDG14 | 8.58 | 32.83 | 10.66 | 42.51 | 5.41 | SDG14 | 1.68 | 2.13 | 2.07 | 2.68 | 2.12 |
| SDG15 | 7.49 | 32.35 | 9.07 | 44.28 | 6.81 | SDG15 | 1.86 | 2.67 | 2.24 | 3.55 | 3.38 |
| SDG16 | 12.55 | 35.55 | 11.47 | 34.59 | 5.85 | SDG16 | 16.48 | 15.51 | 15.00 | 14.65 | 15.36 |
| SDG17 | 10.22 | 30.40 | 13.97 | 38.86 | 6.55 | SDG17 | 5.97 | 5.90 | 8.13 | 7.32 | 7.66 |

*Cognitive relationships*

Although the interlinked nature of SDGs has been stressed, their interactions are "not explicit in the description of the goals" (Griggs et al., 2017). For instance, SDG11, sustainable cities, contains targets related to economic dimensions (e.g., financial and technical assistance for developed countries, expenditure on the conservation on cultural and natural heritage), social dimensions (e.g., number of deaths per disaster and urban population living in slums) or environmental dimensions (e.g., reduce the adverse environmental impact of cities per capita or the proportion of urban solid waste), and the three of them could be conceptually linked to other SDGs like for example SDG6, clean water and sanitation. In our study, to reveal their cognitive relations (measured via citations), a co-citation map has been created. The proximity between SDGs indicates their similarity in terms of co-citation occurrence (i.e., publications from the two SDGs appear often cited together in the same set of publications). The size of the nodes reflects the frequency of SDGs in terms of overall publications, and the thickness of the edges denotes how often these SDGs are co-cited. Figure 6a shows the SDGs map of the M&SDG research. The following clusters of SDGs are identified:

Cluster 1 (red) is formed by SDGs with a strong industrial and energy orientation (SDG6, SDG7 and SDG9), life below water (SDG14), and the environment (SDG15 and SDG14). Cluster 2 (blue) groups SDG1 and SDG2, two of the most important SDGs inheritance of MDGs. SDG1 is directly and indirectly related to all other SDGs, but dependent on SDG2 (International Council for Science, 2015).

Cluster 3 (yellow) includes SDG10, and SDG17, linking the reduction of inequalities and partnership

Cluster 4 (green) is composed by SDGs related with health, urbanization and peace (SDG3; SDG4; SDG 5; SDG11; and SDG16). For instance, SDG11 and SDG3 have a strong connection (link strength of 5,154).

Cluster 5 (purple) is composed only by SDG8, decent work. However, this goal has links with SDG9, industry-related, and SDG11, or SDG3, among others.

Figure 6b depicts the evolution of the SDG in each cluster from the average publication year (2011–2012). The more yellow indicates the more recent the publications. It can be observed how SDGs related with health (SDG3), work and economic development (SDG8), peace (SDG16) and sustainable cities (SDG11) have had research output from earlier years, as compared to the other SDGs. From another perspective, SDGs with a stronger recentness in scientific output include SDG17 (partnership), SDG10 (reduced inequalities), SDG5 (gender equality), and SDG4 (quality education), indicating that awareness of areas related to education or gender are of a more recent nature.

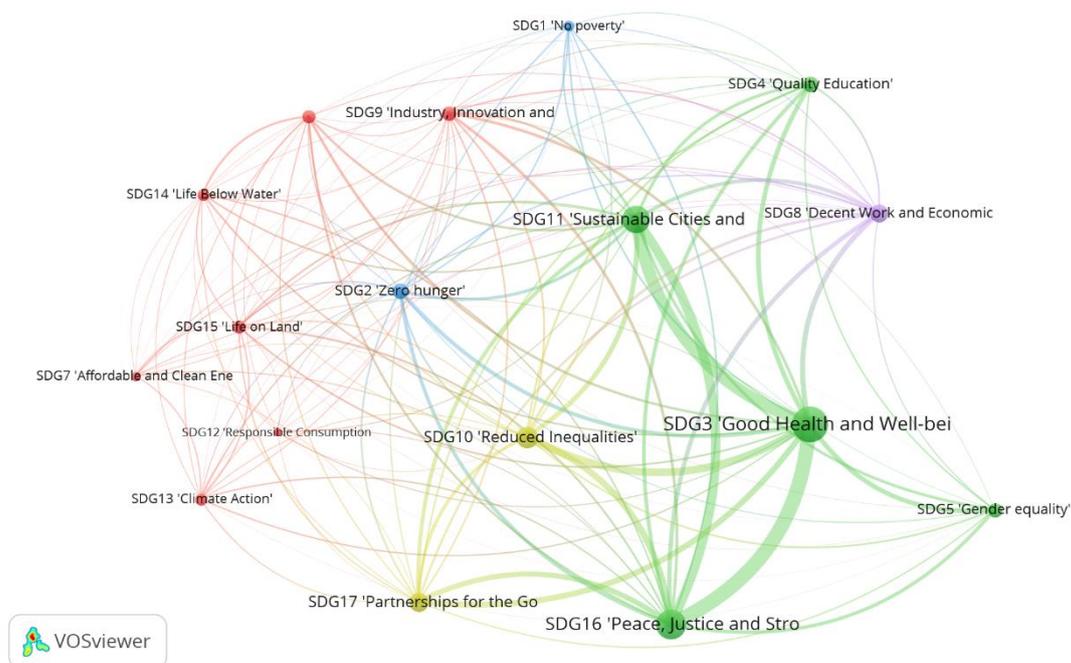

a)

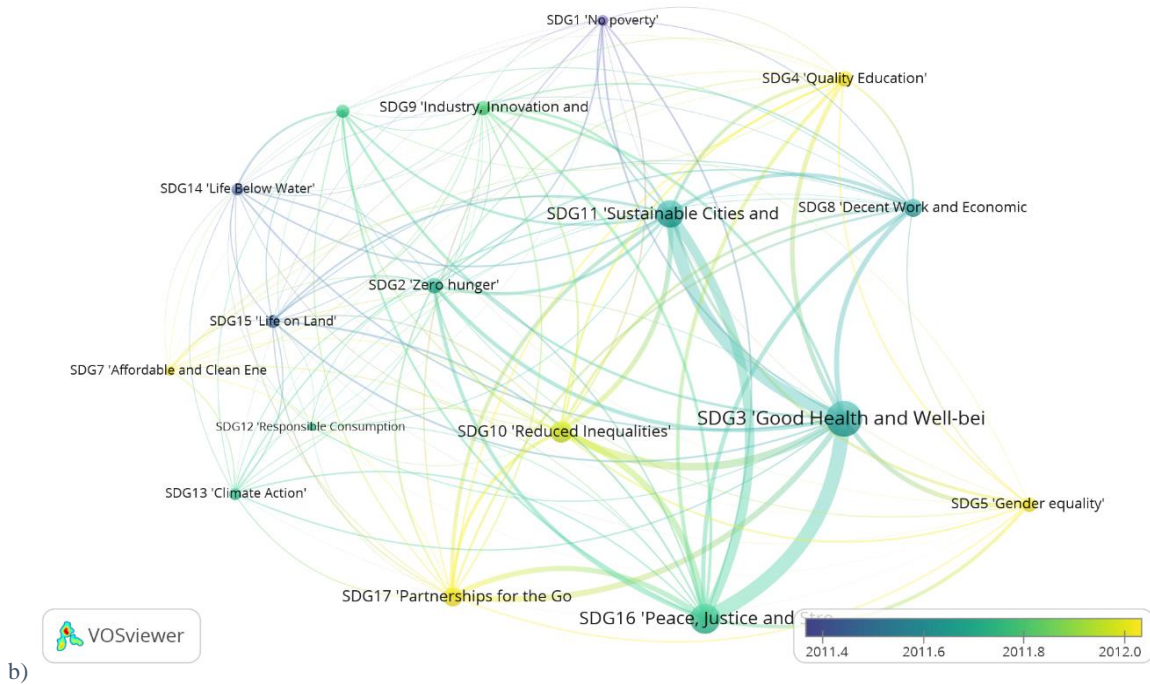

b)

**Figure 6**. Co-citation occurrence map of (a) M&SDGs research and (b) by average publication year.

## 4. Discussion and conclusions

The development of the different SDGs together with the Agenda 2030 have led to the creation of a collective awareness path towards sustainability. One of the main features of SDGs is their increasing relevance not only for policy makers, who are encouraging sustainability-oriented policies, but also for the scientific community as a whole (Kajikawa, 2007). This study presents an empirical analysis of the M&SDGs research and the role of HEIs and RCs in its development. As we stated in the literature review, few studies have focused on analyzing the research output of SDGs, and even fewer have focused on the role of the organizations developing such research. Thus, this paper contributes to the debate around the incorporation of the M&SDGs in the research agenda of HEIs and RCs by providing an overview of output in the area, and by proposing a practical methodology to delineate this area in the WoS database.

*How M&SDGs-research can be bibliometrically delineated and collected?*

A well-delineated methodology is crucial to obtain the research dataset on a specific topic. In this study, we propose an advanced methodology to track and monitor SDGs-related research. The application of our methodology retrieved a total of 25,299 publications, which identifies a much larger set of publications of M&SDGs at HEIs than in similar previous studies (Bizerril, 2018; Veiga-Ávila et al., 2018; Hallinger and Chatpinyakoop, 2019). The recent study by Nakamura (2019) used a very similar methodology as the one presented here, however we identified a larger set of publications related to the M&SDGs realm (4,685 in the core and 25,299 in total in the present study vs. the 2,800 in the core and 10,300 total in Nakamura study). The main reason for this difference is that in the present study the MDGs were also considered, as well as the fact that the CWTS WoS version has a more efficient citation matching algorithm than the one in WoS (Olensky, Schmidt and Van Eck, 2016). The citation-based approach of this study, as well as in Nakamura (2019), offers some advantages in comparison with previous studies that applied keyword-based approaches (Kajikawa et al., 2017; Elsevier, 2015). For instance, it offers a systematic approach that can easily be reproduced and can be applied to any other database that records citation linkages among publications (e.g. Web of Science, Scopus, Microsoft Academic Graph, Dimensions, Crossref Open Citations…), making possible to apply similar approaches in future studies. Another important advantage of this approach is that it focuses on identifying publications that are directly related to M&SDGs, since the selected publications have cited/citing relationship with core literature, thus minimizing the problem of identifying publications with keywords with more circumstantial relationships with M&SDGs topics (just because they carry related keywords), but that are not totally related to them (e.g. publications related to "economic growth", but not in the philosophy underlying the M&SDGs – i.e. sustainable economic growth). Finally, the method developed has the advantage that it captures the research output at the global level, thus providing an international perspective on the discussion around the study of the research activity on

M&SDGs; however in future studies other more local perspectives (e.g. the study of publications in local languages, local publishers) should be also explored.

*How the M&SDGs research by HEIs has been developed over time?*

The results presented in this study suggest that although one may presume that M&SDGs research would have a long tradition with the launch of the MDGs, there is an important concentration of publications in the most recent years, denoting a more recent interest (21.83% in 2000-2014, the MDGs period vs 31.66% from 2015 to 2017 since the launch of the SDGs ). This growing trend is in line with the results by Olawumi and Chan (2018) who observed that the scientific output on SD of 2015 to 2016 represents a 36.27% (vs 21.42% on M&SDGs in the present study). Kajikawa et al. (2007) have stated "around 12,000 papers on sustainability are published annually". There have been previous publications discussing the growth of the scientific production related with sustainability. For example, Pulgarín (2015) argued that the growth of research production in sustainability can be explained by "the impact of human activity on the environment" which "is leading to this area of research [sustainability] being studied from ever more different fields". Olawumi and Chan (2018) consider that this increase could be also linked to "more efforts and resources" being devoted to this topic. For Nučič (2012), the increasing growth of the scientific output could be associated with "sustainability science as a highly interdisciplinary research field". This is in line with Schoolman et al. (2012), who indicated that "sustainability research is more interdisciplinary than other scientific research" (based on the Shannon entropy measure), supporting the suggestion by Nakamura (2019) that SDGs research is based on "transdisciplinary knowledge" between different fields, arguing that "most scientific disciplines are expected to contribute toward sustainability since in sustainability have complex structures, including environmental, technological, societal and economic facets" (Kajikawa et al. (2014)).

In previous studies on M&SDGs the role of HEIs and RC hasn't been specifically analyzed (e.g. in Nakamura, 2019). This study reveals the important participation of HEIS and RCs (85.59% of output is developed by them). Institutions like the London School of Hygiene, the WHO and the John Hopkins University stand out among the most productive institutions. Their predominant role can be explained by their relatively large sizes; however, their AI confirms that these institutions are also highly specialized on this topic too. The London School of Hygiene belongs to the University of London and is specialized in public health and tropical medicine; while the WHO is a specialized agency of the United Nations focused on international public health. The predominant role in output and specialization of the WHO, which is not a HEIs or RCs but a supra-governmental organization that provides statistics for monitoring health-related aspects of the SDGs[12], may be also seen as a sign of the strong social and political relevance of M&SDGs research.

Some other organizations that, although smaller in terms of output, have a high degree of specialization; for example the Stockholm Environment Institute (AI 191.47), the Aga Khan University (AI 141.06) or the International Centre for Diarrhoeal Disease Research, Bangladesh (AI 132.55). This relative importance of small organizations goes in line with Nakamura (2019) results, who suggested that not always the largest institutions 'set the agenda' in M&SDGs research, but that also smaller ones could be key players (e.g., Stockholm University, University of London.)

This study confirms the observation by Yarime et al (2012) of an increasing number of countries engaged in research on sustainability. Our results also resonate with studies like Adomßent (2014), who also stated that the HEIs sustainability research is mostly produced by authors from developed countries such as the USA, the United Kingdom, Australia, or Canada. However, in terms of relative specialization, our study shows that African and Asian countries exhibit a much stronger specialization. A special case is South Africa. This country is the sixth country in number of M&SDGs publications, and one of its universities (i.e. University of Cape Town) is the most prolific African institution in M&SDGs research. This 6[th] position contrasts with the 113[th] position (out of 162 countries) in the SDG index (Stiftung and SDSN, 2019), which suggests that countries with stronger difficulties in improving their overall achievement of SDGs (as measured by the SDG index), may be prone to invest more research on SDGs. This strong relevance of SDGs research in the South African research context is somehow reinforced by the fact that South Africa is also a 'topic' by itself in M&SDGs research, since the name of the country appears as a node in the term co-occurrence map.

*What are the thematic relationships between SDGs?*

The SDGs more frequently addressed by HEIs and RCs are SDG3 (76.93% of the publications), SDG16 (56.19%), SDG11 (45.98%) and SDG10 (29.47%), which is in line with the higher percentage of overall HEIs and RCs involved on this research (Table A.3.). Our results contrast with the results obtained by (Salvia et al., 2019), who highlighted other SDGs to experts according to their experience and research area: SDG 13 (41%), SDG 11 (33%) and SDG 4 (29%). Salvia et al (2019) results were obtained by surveying expert across continents, therefore the methodology applied differs substantially from the one in this study. However, we still find a similar degree of activity for SDG11 in both studies.

---

[12] Information available at: https://www.who.int/sdg/en/

As mentioned above, SDG3 (Good health) is the most researched SDG identified in our study, which is not a surprise since this SDG is centrally positioned in the Agenda 2030 (WHO, 2006) and has a central role in the achievement of SD (Pettigrew et al., 2015). The major efforts on this specific SDG have been made to reduce mortality across population groups (e.g. 'the poor' or 'women and children') (Buse and Hawkes, 2015). This central importance of "good health" is in line with a large number of health-related keywords obtained in this study. Besides, topics related to maternal or child mortality appear with a stronger recentness, which coincides with the SDGs launch.

Regarding the interconnection of SDGs, according to Nilsson (2016), SDGs are more interconnected among themselves than its predecessors, the MDGs. This idea of SDGs interconnecting among themselves is supported by their consideration as "enablers for integration", which means that the internal structures of the different SDGs is conceived to fit across more different SDGs (Le Blanc, 2015), thus enabling their own integration and interconnection. This integrative and interconnected property of SDGs is observed in this study, since all have connections among them, being particularly remarkable the connections between the following pairs: SDG16-SDG13; SDG3-SDG11; SDG16-SDG11. Moreover, the linkage between the pairs responds to complementary relationships. For instance, SDG10 and SDG17, linking the reduction of inequalities and partnerships, could be understood as that data should be collected of all groups of population and analysed in the disaggregated form to ensure targets are being met for everyone (International Council for Science, 2015).

*Final remarks*

This paper has provided an extensive analysis of M&SDGs research over time and contributes to understand their trajectory This paper confirms the central role of HEIs and RCs in M&SDGs research This study provides a novel contribution to the bibliometric analysis on the SDGs research output, although it also presents some limitations that must be observed when generalizing its findings. For instance, the methodology proposed may not necessarily capture the whole picture of research *related to* M&SDGs. The use of just direct citations related to a *core* set of publications may be insufficient at times, since many publications genuinely linked to M&SDGs research may be more distanced in their citation relationship with the *core* set. Thus, future methodological improvements should contemplate the possibility of characterizing not only the directly cited/citing publications related to M&SDGs, but also other citation layers (e.g. $2^{nd}$, $3^{rd}$ or more) in the expansion of the *core* set of publications, thus moving towards a more fluid approach (in which a larger set of publications may be considered in their citation proximity to the *core* set) in contrast with the binary approach considered in this study. Moreover, future studies on the topic might be complemented by means of qualitative research methods to uncover motivations and drivers for research on SDGs in different contexts. The combination of bibliometric indicators with other monitoring indicators (e.g. the SDG index) should also be considered, in other to provide more advanced analysis about the relationship between the research efforts of countries (as done in this study) and their success in their actual achievement of the specific SDGs, thus providing a more holistic perspective on how research can complement and support the consecution of the Agenda 2030 of SDGs.

# 5. References

Adomßent, M., Fischer, D., Godemann, J., Herzig, C., Otte, I., Rieckmann, M., & Jana Timma, J. (2014). Emerging areas in research on higher education for sustainable development e management education, sustainable consumption and perspectives from Central and Eastern Europe. *Journal of Cleaner Production, 62*, 1-7. doi:

Alejandro-Cruz, J. S., Rio-Belver, R. M., Almanza-Arjona, Y. C., & Rodriguez-Andara, A. (2019). Towards a Science Map on Sustainability in Higher Education. *Sustainability*, *11*(13), 3521.

Bautista-Puig, N., & Mauleón, E. (2019a). Unveiling the path towards sustainability: is there a research interest on sustainable goals? *International Conference on Scientometrics & Informetrics*, Rome, Italy, 17. (pp. 2770-2771).

Bautista, N. (2019b): *SDGs ontology*. figshare. Dataset. https://doi.org/10.6084/m9.figshare.11106113.v1

Bizerril, M., Rosa, M. J., Carvalho, T., & Pedrosa, J. (2018). Sustainability in higher education: A review of contributions from Portuguese Speaking Countries. *Journal of Cleaner Production, 171*, 600-612. doi:10.1016/j.jclepro.2017.10.048

Buse, K., & Hawkes, S. (2015). Health in the sustainable development goals: Ready for a paradigm shift? *Globalization and Health*, 11(1), 1–8. https://doi.org/10.1186/s12992-015-0098-8

Caeiro, S., Azeiteiro, U. M., Filho, W. L., & Jabbour, C. (Eds.). (2013). *Sustainability Assessment Tools in Higher Education Institutions: Mapping Trends and Good Practices Around the World* (pp. 65–78). Cham: Springer International Publishing. doi.org/10.1007/978-3-319-02375-5_4

Caiado, R. G. G., Filho, W. L., Quelhas, O. L. G., Nascimento, D. L. d. M., & Ávila, L. V. (2018). A literature-based review on potentials and constraints in the implementation of the sustainable development goals. *Journal Cleaner Production, 198*, 1276-1288. doi: 10.1016/j.jclepro.2018.07.102

Easterly, W. (2009). How the millennium development goals are unfair to Africa. World development, 37(1), 26-35.

Elsevier Research Intelligence (2015). *Sustainability Science in a Global Landscape*. https://doi.org/10.1016/j.aqpro.2013.07.003

Frame, J. D. (1977). Mainstream research in Latin America and the Caribbean. *Interciencia*, 2(3), 143-148.

Fukuda-Parr, S. (2016). From the Millennium Development Goals to the Sustainable Development Goals: shifts in purpose, concept, and politics of global goal setting for development. *Gender & Development, 24*(1), 43-52.

Griggs, D. J., Nilsson, M., Stevance, A., & McCollum, D. (2017). *A guide to SDG interactions: from science to implementation*. International Council for Science, Paris.

Hallinger, P., & Chatpinyakoop, C. (2019). A bibliometric review of research on higher education for sustainable development, 1998-2018. *Sustainability (Switzerland), 11*(8). doi.org/10.3390/su11082401

Hák, T., Janoušková, S., & Moldan, B. (2016). Sustainable Development Goals: A need for relevant indicators. *Ecological Indicators, 60*, 565–573.

Hassan, S. U., Haddawy, P., & Zhu, J. (2014). A bibliometric study of the world's research activity in sustainable development and its sub-areas using scientific literature. Scientometrics, 99(2), 549–579.

Hesselbarth, C., & Schaltegger, S. (2014). Educating change agents for sustainability–learnings from the first sustainability management master of business administration. Journal of Cleaner Production, 62, 24–36.

International Council for Science (2015): Review of the Sustainable Development Goals: The Science Perspective. Paris: International Council for Science (ICSU).

International Human Rights Instruments. (2008). *Report on Indicators for Promoting and Monitoring the Implementation of Human Rights.* Geneva: United Nations.

Kajikawa, Y., Ohno, J., Takeda, Y., Matsushima, K., & Komiyama, H. (2007). Creating an academic landscape of sustainability science: an analysis of the citation network. *Sustainability Science, 2*(2), 221.

Kajikawa, Y., Tacoa, F., & Yamaguchi, K. (2014). Sustainability science: the changing landscape of sustainability research. Sustainability science, 9(4), 431-438.

Ki-Moon, B. (2013). The millennium development goals report 2013. *United Nations Publications*.

Kroll, C. (2015). Sustainable development goals: Are the rich countries ready? SDG index report 2015. Available at: https://www.sdgindex.org/reports/sdg-index-report-2015/ accessed on 10 November 2019

Kleinberg, J. (2003). Bursty and hierarchical structure in streams. Data Mining and Knowledge Discovery, 7(4), 373-397

Leal Filho, W., Wu, Y.-C. J., Brandli, L. L., Avila, L. V., Azeiteiro, U. A., Caeiro, S., & Madruga, L. R. d. R. G. (2017). Identifying and overcoming obstacles to the implementation of sustainable development at universities. *Journal of Integrative Environmental Sciences, 14*(1), 93-108. doi:DOI: 10.1080/1943815X.2017.1362007

Leal Filho, W., Pallant, E., Enete, A., Richter, B., & Brandli, L. L. (2018). Planning and implementing sustainability in higher education institutions: an overview of the difficulties and potentials. *International Journal of Sustainable Development and World Ecology*. doi:https://doi.org/10.1080/13504509.2018.1461707

Le Blanc, D. (2015). Towards integration at last? The sustainable development goals as a network of targets. *Sustainable Development, 23*(3), 176-187.

Lozano, R., Ceulemans, K., Alonso-Almeida, M., Huisingh, D., Lozano, F. J., Waas, T., … Hugé, J. (2015). A review of commitment and implementation of sustainable development in higher education: Results from a worldwide survey. *Journal of Cleaner Production, 108*, 1–18. https://doi.org/10.1016/j.jclepro.2014.09.048

Nakamura, M., Pendlebury, D., Schnell, J., & Szomszor, M. (2019). Navigating the Structure of Research on Sustainable Development Goals. Policy, 11, 12.

Nilsson, M., Griggs, D., & Visbeck, M. (2016). Policy: map the interactions between Sustainable Development Goals. *Nature News, 534*(7607), 320.

Nučič, M. (2012). Is sustainability science becoming more interdisciplinary over time?. Acta geographica Slovenica, 52(1), 215-236.

Olawumi, T. O., & Chan, D. W. (2018). A scientometric review of global research on sustainability and sustainable development. Journal of Cleaner Production, 183, 231-250.

Olawumi, T. O., & Chan, D. W. M. (2018). A scientometric review of global research on sustainability and sustainable development. *Journal Clean Production, 183*, 231-250. doi:https://doi.org/10.1016/j.jclepro.2018.02.162

Olensky, M., Schmidt, M., & Van Eck, N.J. (2016). Evaluation of the Citation Matching Algorithms of CWTS and iFQ in Comparison to the Web of Science. *Journal of the Association for Information Science and Technology*, 67(10), 2550-2564. doi:10.1002/asi.23590

Pettigrew, L. M., De Maeseneer, J., Anderson, M. I. P., Essuman, A., Kidd, M. R., & Haines, A. (2015). Primary health care and the Sustainable Development Goals. The Lancet, 386(10009), 2119-2121.

Pulgarin, A., Eklund, P., Garrote, R., & Escalona-Fernandez, M. I. (2015). Evolution and structure of "sustainable development": A bibliometric study. Brazilian journal of Information Science: Research Trends, 9(1), 24.

Ramírez Ríos, J. F., Alzate Ibáñez, A. M., & Montenegro Riaño, D. F. (2016). Los discursos de la sostenibilidad: análisis de tendencias conceptuales a partir de mediciones bibliométricas. *Questionar: Investigación Específica, 4*(1), 82-96.

Salvia, A. L., Leal Filho, W., Brandli, L. L., & Griebelera, J. S. (2019). Assessing research trends related to Sustainable Development Goals: local and global issues. *Journal Clean Production, 208*, 841-849. doi:https://doi.org/10.1016/j.jclepro.2018.09.242

Schoolman, E. D., Guest, J. S., Bush, K. F., & Bell, A. R. (2012). How interdisciplinary is sustainability research? Analyzing the structure of an emerging scientific field. *Sustainability Science, 7*(1), 67-80.

Stifstung; Sustainable Develoment Solutions Network (SDSN) (2019). Sustainable development report 2019. Transformations to achieve the Sustainable Development Goals. Includes the SDG Index and Dashboards. Available at: https://s3.amazonaws.com/sustainabledevelopment.report/2019/2019_sustainable_development_report.pdf

Stiftung; Sustainable Develoment Solutions Network (SDSN) (2019). Sustainable Development Solutions Network. SDG Index and Dashboards Report 2019. Global responsibilities: implementing the goals. 2019.

Tatalović and Antony (2010). What Has It Done For The Millennium Development Goals?' Available at:
http://www.scidev.net/global/health/feature/science-what-has-it-done-for-the-millennium-development-goals--1.html

United Nations (1987). *Our common future: Report of the 1987 world commission on environment and development.* Oslo: United Nations.

United Nations (2015a). Millennium Development Goals: 2015 Progress Chart. Avalailble at: https://www.un.org/millenniumgoals/2015_MDG_Report/pdf/MDG%202015%20PC%20final.pdf

UNESCO. (2017). *Educação para os Objetivos de Desenvolvimento Sustentável: objetivos de aprendizagem.* In.

United Nations. (2015b). Transforming Our World: The 2030 Agenda for Sustainable Development (A/RES/70/1). Retrieved from https://undocs.org/A/RES/70/1

Veiga Ávila, L., Rossato Facco, A. L., Bento, M. H. dos S., Arigony, M. M., Obregon, S. L., & Trevisan, M. (2018). Sustainability and education for sustainability: An analysis of publications from the last decade. *Environmental Quality Management, 27*(3), 107–118. Doi:10.1002/tqem.21537

Van Eck, N. J., & Waltman, L. (2017). Citation-based clustering of publications using CitNetExplorer and VOSviewer. *Scientometrics, 111*(2), 1053-1070.

Waltman, L., Calero-Medina, C., Kosten, J., Noyons, E. C., Tijssen, R. J., van Eck, N. J., van Leeuwen T., & Wouters, P. (2012). The Leiden Ranking 2011/2012: Data collection, indicators, and interpretation. *Journal of the American society for information science and technology*, *63*(12), 2419-2432.

Yarime, M., Takeda, Y., & Kajikawa, Y. (2010). Towards institutional analysis of sustainability science: a quantitative examination of the patterns of research collaboration. Sustainability Science, 5(1), 115.

Waas, T., Verbruggen, A., & Wright, T. (2010). University research for sustainable development: Definition and characteristics explored. Journal of Cleaner Production, 18(7), 629–636.

World Health Organization. (2016). World health statistics 2016: monitoring health for the SDGs sustainable development goals. World Health Organization.

Wuelser, G., & Pohl, C. (2016). How researchers frame scientific contributions to sustainable development: a typology based on grounded theory. *Sustainable Science, 11*(789–800). doi:DOI 10.1007/s11625-016-0363-7

# Appendix:

**Definitions:**

- Higher Education Institutes (HEIs): include traditional scientific universities and professional-oriented institutions, which are called universities of applied sciences or polytechnics.
- Research centre (RC): a "formally structured unit within the university, other than a department or a school, established with the purpose of advancing scholarly activity primarily through collaborative research, research training, research dissemination, or creative endeavours"[13].

**Eq (A.1.).** The AI was defined by the following equation:

$$AI = \frac{\sum_{i=1}^{n} {}_{Ac}^{M\&SDG}p_i}{\sum_{i=1}^{n} {}_{WoS}^{M\&SDG}p_i} \Big/ \frac{\sum_{i=1}^{n} {}_{Ac}^{All}p_i}{\sum_{i=1}^{n} {}_{WoS}^{All}p_i} \quad (i = 2000, \ldots, 2017)$$

where:

- ${}_{Ac}^{M\&SDG}p_i$ is the total output of an actor ($Ac$) on the M&SDGs topic in the Web of Science;
- ${}_{WoS}^{M\&SDG}p_i$ is the total output in Web of Science in M&SDGs (n= 25,299)
- $\sum_{i=1}^{n} {}_{Ac}^{All}p_i$ is the total output of an actor ($Ac$) in WoS in the same period.
- $\sum_{i=1}^{n} {}_{WoS}^{All}p_i$ is the total number of documents in Web of Science during the period (n=30,549,291)

**Table A.1.** 6-year Period Evolution of the Top 20 Organizations Participating in Sustainability Goals Research

| 2000-2005 | | | 2006-2011 | | | 2012-2017 | | |
|---|---|---|---|---|---|---|---|---|
| Organization | No. Docs | % | Organization | No. Docs | % | Organization | No. Docs | % |
| World Health Organization | 292 | 12,53 | London School of Hygiene & Tropical Medicine | 682 | 10,22 | London School of Hygiene & Tropical Medicine | 1009 | 7,98 |
| London School of Hygiene & Tropical Medicine | 272 | 11,67 | World Health Organization | 580 | 8,69 | World Health Organization | 803 | 6,35 |
| Johns Hopkins University | 157 | 6,74 | Johns Hopkins University | 439 | 6,58 | Johns Hopkins University | 728 | 5,75 |
| Harvard University | 117 | 5,02 | Harvard University | 318 | 4,77 | Harvard University | 644 | 5,09 |
| University of Oxford | 69 | 2,96 | University of Oxford | 245 | 3,67 | University of Washington, Seattle | 479 | 3,79 |
| Columbia University | 67 | 2,88 | University College London | 209 | 3,13 | University of Oxford | 430 | 3,40 |
| University of California, Berkeley | 52 | 2,23 | Columbia University | 203 | 3,04 | University College London | 385 | 3,04 |
| University College London | 47 | 2,02 | Imperial College London | 192 | 2,88 | Columbia University | 380 | 3,00 |
| University of Liverpool | 44 | 1,89 | University of Cape Town | 178 | 2,67 | University of the Witwatersrand | 360 | 2,85 |
| Imperial College London | 44 | 1,89 | University of Liverpool | 165 | 2,47 | University of Cape Town | 355 | 2,81 |
| Cornell University | 42 | 1,80 | University of Washington, Seattle | 142 | 2,13 | Imperial College London | 345 | 2,73 |
| University of Cape Town | 41 | 1,76 | International Centre for Diarrhoeal | 131 | 1,96 | University of Toronto | 331 | 2,62 |

---

[13] Definition extracted from the following link: https://www.ufv.ca/media/assets/secretariat/policies/Research-Centres-and-Institutes-(211).pdf accessed on 5 February 2020

| | | | | | | | | |
|---|---|---|---|---|---|---|---|---|
| | | | | Disease Research, Bangladesh | | | | |
| Swiss Tropical & Public Health Institute | 40 | 1,72 | Aga Khan University | 129 | 1,93 | University of Queensland | 300 | 2,37 |
| University of East Anglia | 39 | 1,67 | University of California, San Francisco | 128 | 1,92 | University of North Carolina, Chapel Hill | 298 | 2,36 |
| University of North Carolina, Chapel Hill | 38 | 1,63 | University of California, Berkeley | 126 | 1,89 | University of Melbourne | 297 | 2,35 |
| University of Sussex | 36 | 1,55 | University of the Witwatersrand | 120 | 1,80 | Aga Khan University | 297 | 2,35 |
| Yale University | 36 | 1,55 | University of Queensland | 112 | 1,68 | University of Liverpool | 296 | 2,34 |
| University of Toronto | 34 | 1,46 | Swiss Tropical & Public Health Institute | 109 | 1,63 | Makerere University | 265 | 2,09 |
| International Centre for Diarrhoeal Disease Research, Bangladesh | 34 | 1,46 | Makerere University | 107 | 1,60 | Emory University | 251 | 1,98 |
| Federal University of Pelotas | 33 | 1,42 | Emory University | 102 | 1,53 | Karolinska Institute | 244 | 1,93 |
| Total docs. | 2,330 | | | 6,671 | | | 12,652 | |

**Table A.2**. Top 60 Keywords with the Strongest Citation Bursts Sorted by Opening Year

| Keywords | Strength | Begin | End | 2000 - 2017 |
|---|---|---|---|---|
| infant mortality | 41.0873 | 2000 | 2008 | |
| income | 15.9272 | 2000 | 2003 | |
| low birth weight | 11.4649 | 2000 | 2003 | |
| trial | 22.5934 | 2000 | 2004 | |
| inequality | 48.7206 | 2000 | 2008 | |
| nutrition | 6.0144 | 2000 | 2001 | |
| globalization | 16.9567 | 2000 | 2006 | |
| health care | 18.6416 | 2000 | 2006 | |
| antenatal care | 9.5531 | 2000 | 2003 | |
| transmission | 40.9383 | 2000 | 2007 | |
| aid | 31.8646 | 2000 | 2007 | |
| cost | 17.542 | 2000 | 2006 | |
| human immunodeficiency virus | 6.6829 | 2000 | 2001 | |
| economic growth | 8.02 | 2000 | 2001 | |
| prenatal care | 11.092 | 2000 | 2004 | |
| rural Bangladesh | 5.346 | 2000 | 2001 | |
| anemia | 6.0144 | 2000 | 2001 | |
| randomized trial | 8.1832 | 2001 | 2004 | |
| malaria | 14.4421 | 2001 | 2006 | |
| safe motherhood | 8.1832 | 2001 | 2004 | |
| tuberculosis | 66.5183 | 2001 | 2011 | |
| fertility | 12.3561 | 2001 | 2003 | |
| growth | 30.4484 | 2001 | 2008 | |
| infection | 20.8902 | 2002 | 2008 | |
| Uganda | 9.8743 | 2002 | 2003 | |
| cost effectiveness | 17.0745 | 2002 | 2006 | |
| gender | 9.6188 | 2002 | 2004 | |
| Tanzania | 8.7135 | 2002 | 2004 | |
| malnutrition | 33.955 | 2003 | 2009 | |
| education | 10.544 | 2003 | 2004 | |
| public health | 42.1686 | 2004 | 2012 | |
| morbidity | 53.5406 | 2006 | 2011 | |
| poverty | 20.0308 | 2006 | 2009 | |
| randomized controlled trial | 52.8332 | 2006 | 2013 | |
| Kenya | 18.7307 | 2007 | 2008 | |
| strategy | 45.8455 | 2007 | 2012 | |
| population | 34.6716 | 2007 | 2011 | |
| neonatal mortality | 11.2422 | 2007 | 2008 | |
| model | 22.6023 | 2008 | 2010 | |
| HIV | 19.8909 | 2009 | 2013 | |
| sustainability | 6.7965 | 2009 | 2011 | |
| survival | 12.697 | 2010 | 2013 | |
| challenge | 9.4519 | 2011 | 2012 | |
| child mortality | 49.1852 | 2011 | 2014 | |
| United States | 8.4907 | 2011 | 2012 | |
| prevention | 16.4552 | 2012 | 2013 | |
| systematic analysis | 60.5619 | 2013 | 2017 | |
| burden | 27.8308 | 2013 | 2017 | |
| millennium development goal | 28.3765 | 2013 | 2014 | |
| community | 8.0114 | 2013 | 2014 | |
| equity | 33.994 | 2013 | 2015 | |
| trend | 30.5064 | 2013 | 2017 | |
| epidemiology | 5.7865 | 2013 | 2014 | |
| maternal health | 64.9803 | 2014 | 2017 | |
| access | 57.4133 | 2014 | 2017 | |
| middle income country | 75.1318 | 2014 | 2017 | |
| delivery | 36.3786 | 2014 | 2017 | |
| infant | 14.7382 | 2014 | 2015 | |
| outcome | 38.0628 | 2015 | 2017 | |
| newborn | 16.6561 | 2015 | 2017 | |

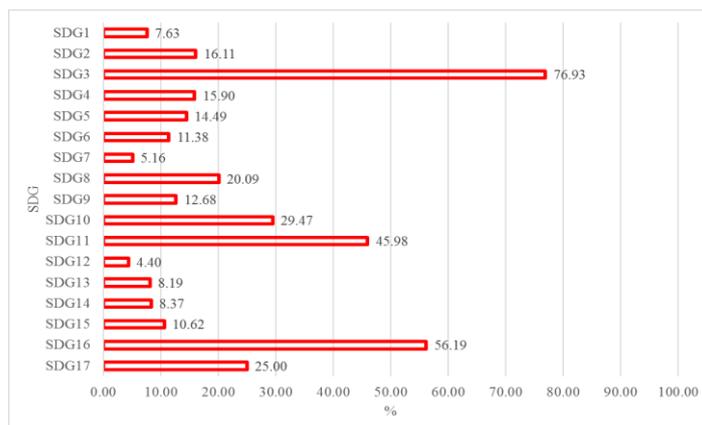

**Figure A.1.** Percentage of documents classified by each SDG (n=20,749)

**Table A.3.** Approach of the 17 SDGs by the Institutions Analysed

| Goal | Number of institutions involved | % |
|---|---|---|
| SDG1 | 724 | 36.79 |
| SDG2 | 1,045 | 53.10 |
| SDG3 | 1,670 | 84.86 |
| SDG4 | 1,033 | 52.49 |
| SDG5 | 891 | 45.27 |
| SDG6 | 937 | 47.61 |
| SDG7 | 678 | 34.45 |
| SDG8 | 1,161 | 58.99 |
| SDG9 | 1,101 | 55.95 |
| SDG10 | 1,380 | 70.12 |
| SDG11 | 1,505 | 76.47 |
| SDG12 | 624 | 31.71 |
| SDG13 | 973 | 49.44 |
| SDG14 | 890 | 45.22 |
| SDG15 | 1,006 | 51.12 |
| SDG16 | 1,532 | 77.85 |
| SDG17 | 1,385 | 70.38 |
| Total institutions | 1,968 | |